\documentclass[twocolumn,prd,amsmath,superscriptaddress,nofootinbib,floatfix]{revtex4}

\usepackage{amsmath}
\usepackage{bm}
\usepackage{graphics}

\def\gsim{\;\rlap{\lower 2.5pt
 \hbox{$\sim$}}\raise 1.5pt\hbox{$>$}\;}
\def\lsim{\;\rlap{\lower 2.5pt
   \hbox{$\sim$}}\raise 1.5pt\hbox{$<$}\;}

\def\be{\begin{equation}}
\def\ee{\end{equation}}
\def\bea{\begin{eqnarray}}
\def\eea{\end{eqnarray}}

\begin{document}

\title{Constraining Dark Energy by Combining Cluster Counts and\\Shear--Shear
Correlations in a Weak Lensing Survey}

\author{Wenjuan Fang}
\affiliation{Department of Physics, Columbia University, New York,
NY 10027, USA}
\author{Zolt\'{a}n Haiman}
\affiliation{Department of Astronomy, Columbia University, New York,
NY 10027, USA}

\date{\today}

\begin{abstract}
  We study the potential of a large future weak--lensing survey to
  constrain dark energy properties by using both the number counts of
  detected galaxy clusters (sensitive primarily to density
  fluctuations on small scales) and tomographic shear--shear
  correlations (restricted to large scales). We use the Fisher matrix
  formalism, assume a flat universe and parameterize the equation of
  state of dark energy by $w(a)=w_0+w_a(1-a)$, to forecast the
  expected statistical errors from either observable, and from their
  combination.  We show that the covariance between these two
  observables is small, and argue that therefore they can be regarded
  as independent constraints.  We find that when the number counts and
  the shear-shear correlations (on angular scales $\ell\leq 1000$) are
  combined, an LSST (Large Synoptic Survey Telescope)--like survey can
  yield statistical errors on $\Omega_{\rm DE}, w_0, w_a$ as tight as
  0.003, 0.03, 0.1.  These values are a factor of $2-25$ better than
  using either observable alone. The results are also about a factor
  of two better than those from combining number counts of galaxy
  clusters and their power spectrum.
\end{abstract}

\maketitle

\section{Introduction}

The existence of dark energy is strongly indicated by the relatively
dim appearance of distant supernovae~\cite{SP99,AR98}, by the
shortfall of the matter density to make the universe spatially
flat~(e.g.\cite{D&K00}), and by recent, accurate measurements of
cosmic microwave background (CMB) anisotropy~\cite{CLB03,WMAP3}.  This
newly discovered form of energy constitutes about 2/3 of the total
energy density of the universe, and is known to have negative pressure
and a nearly uniform spatial distribution. Several competing
theoretical models have been proposed to explain dark energy (see,
e.g., \cite{H&T01} for a list of references).  While current
observational data cannot distinguish among these proposals, it is
hoped that future observations, which will reach higher precision, can
constrain models, and clarify the nature of dark energy.

In this paper, we explore one of the methods to constrain dark energy
parameters in the future: using the weak gravitational lensing (WL)
distortion of distant galaxies in a large survey of the sky.  The
light from distant galaxies is deflected by the foreground
gravitational field, causing small but statistically coherent
distortions in the observed shapes of the galaxies. This so--called
weak-lensing shear signal can be observed, in principle, for a large
number of galaxies, and used to infer the foreground gravitational
field or almost equivalently, the mass distribution (see recent
reviews by, e.g., \cite{PS06, VW&M03}).

Several previous studies have examined the constraints that can be
placed on dark energy from large weak--lensing surveys.  The most
commonly proposed method is to utilize statistical properties of the
two--dimensional shear field directly, such as its power spectrum (or,
equivalently, the shear--shear correlation function).  With additional
photometric redshift information, which future wide--field
multi--color surveys are expected to provide, the background galaxies
can be grouped into different redshift bins.  Statistics done within
each bin and among different bins can recover some of the information
from the third (radial) dimension.  Such a ``tomographic'' study of
the foreground density fluctuations provides an additional handle on
dark energy, through the effect of dark energy on the recent
($0<z\lsim 1$) growth of dark matter perturbations and the expansion
history of the universe
(e.g.~\cite{WH99,WH01,DH01,AR04,A&D03,T&J04,S&K04}).  In particular,
Song \& Knox~\cite{S&K04} have evaluated the statistical constraints
from tomographic shear--shear correlations that are expected to be
available from the Large Synoptic Survey Telescope (LSST).

A complementary method is to utilize statistical properties of the
{\em peaks} of the shear field.  This method, however, does not lend
itself to straightforward mathematical analysis, and has been
relatively much less well explored~(e.g.\cite{BJ&VW00,SW06}) On the
other hand, there is an increasingly better correspondence between
higher--$\sigma$ shear peaks and discrete, massive virialized objects
--galaxy clusters-- in the foreground~\cite{H&T&Y03,H&S05}. To the
extent that this correspondence can be quantified ab--initio in
numerical simulations, one can use the shear--selected cluster sample,
including their abundance evolution and power spectrum, to constrain
dark energy properties (e.g.~\cite{SW04,M&B06}).  In general, the
abundance evolution of galaxy clusters can place strong constraints on
dark energy parameters, because it is exponentially sensitive to the
growth rate of matter fluctuations. Several studies have explored
constraints expected from future surveys that would detect clusters
through their X--ray emission or the Sunyaev-Zeldovich effect
~(e.g.~\cite{W&S98,H&M&H01,JW02,JW03}.  An attractive feature of
utilizing weak lensing signal to detect clusters is that it directly
probes the total mass of the cluster.  In particular, Wang et
al.~\cite{SW04} have evaluated the statistical constraints from the
galaxy cluster sample that is expected to be available from LSST.

Once the data from a wide--field WL survey, such as LSST~\cite{lsst},
or a smaller pre--cursor mission, such as Pan-Starr~\cite{panstarr}
has been collected, it will be logical to try to extract information
on dark energy properties from {\it both} the shear--shear
correlations and cluster abundance, since both pieces of information
will be available in the same data--set.  {\em The goal of the present
paper is to quantify the improvement on dark energy constraints when
these two pieces of information are combined.}

Ideally, one would pose more ambitious questions, such as: what is the
maximum information one can obtain on dark energy parameters, given
the effect of dark energy on the full non--linear shear field?  In
particular, it is not clear whether either statistic (shear--shear
correlations or cluster counts), in fact, captures a significant
fraction of the available information.  Nonetheless, in this paper, we
contend ourselves to answering the much simpler question
above. Furthermore, for simplicity, we will use the shear-shear
correlation function only on large angular scales ($\ell\leq
1000$). On large scales, where density fluctuations are in the linear
regime, the correlation function contains all the information about
the density field. On smaller scales, non--linear evolution introduces
significant non--Gaussianity.  We will argue below that once cluster
counts are taken into account, the small--scale shear--shear
correlations offer no significant additional information.  In this
paper, we focus on evaluating constraints for a ground--based survey,
such as LSST.  We parameterize dark energy by its bulk equation of
state $w=\langle P\rangle/\langle\rho\rangle$~\cite{T&W97}, and allow
$w$ to evolve linearly with the cosmic scale--factor $a$ as
$w(a)=w_0+w_a(1-a)$.

This paper is organized as follows. In \S~\ref{sec:II}, we describe
our calculational methods, which closely follow previous Fisher matrix
analyses, but include an additional discussion of covariance.  Our
results are presented in \S~\ref{sec:III}, and discussed, along with
various caveats, in \S~\ref{sec:IV}.  Finally, in \S~\ref{sec:V}, we
offer our conclusions and summarize the implications of this work.

\section{Calculational Methods}
\label{sec:II}

\subsection{Cluster Number Counts}

We closely follow ref.~\cite{SW04}, and consider a sample of
shear--selected clusters with specifications motivated by LSST. In
particular, we assume the sample covers the redshift range of
$0.1\leq z \leq 1.4$, which is divided into 26 redshift bins of
equal size $\Delta z=0.05$. The expected number of clusters in the
$i^{\rm th}$ redshift bin is calculated as
\begin{equation}
\bar{N_i}=\Delta \Omega \Delta
z\frac{d^2V}{dzd\Omega}(z_i)\int_{M_{\rm
min}(z_i)}^{\infty}\frac{dn}{dM}(M,z_i)dM \label{eqn:Ni}
\end{equation}
where $z_i$ is the central redshift for the $i^{\rm th}$ redshift
bin, $\Delta \Omega$ is the solid angle covered by the survey, which
for LSST, we take to be $18,000$ $\rm deg^2$, $d^2V/dzd\Omega$ is
the comoving volume element, $M_{\rm min}$ is the detection
threshold mass for clusters, and $\frac{dn}{dM}$ is the cluster mass
function.

We use the fitting formula given by Jenkins {\it et al.}
~\cite{AJ01} for the cluster mass function,
\begin{multline}
\frac{dn}{dM}(M,z)=0.301\frac{\rho_{m}}{M}\frac{d\ln\sigma^{-1}(M,z)}{dM} \\
\times \exp(-|\ln\sigma^{-1}(M,z)+0.64|^{3.82})
\end{multline}
where $\rho_m$ is the present--day matter density, $\sigma(M,z)$ is
the amplitude of the linear matter fluctuations at redshift $z$,
smoothed by a top hat window function whose scale is such that the
enclosed mass at the mean density $\rho_m$ is $M$. This formula is
based on the identification of dark matter halos as spherical regions
that have a mean overdensity of 180 with respect to the background
matter density at the time of identification.

In determining the mass threshold for detecting a cluster at redshift
z, $M_{\rm min}(z)$, we follow NFW~\cite{NFW97}, and model the density
profile of galaxy clusters with the self--similar function
\begin{equation}
\rho_{\rm NFW}(r)=\frac{\rho_s}{(r/r_s)[1+(r/r_s)]^2},
\end{equation}
where $r$ is the radius from the cluster center, and $r_s, \rho_s$
are some characteristic radius and density. In ~\cite{NFW97}, the
density profile is truncated at $r_{200}$, inside which the mean
overdensity with respect to the critical density of the universe at
redshift $z$ is 200. The outer radius $r_{200}$ is parameterized by
the concentration parameter, defined as
\begin{equation}
c_{\rm NFW}=\frac{r_{200}}{r_s}
\end{equation}
and taken to be a constant $c_{\rm NFW}=5$ in this paper. With the
definition of the cluster mass $M_{200}$ as the mass enclosed within
$r_{200}$, the cluster's structure ($\rho_s, r_s, r_{200}$) is fully
determined by $M_{200}$ and $z$.

We then follow Hamana {\it et al.} ~\cite{H&T&Y03}, and use a Gaussian
window function to smooth the convergence field induced by the
cluster,
\begin{equation}
\kappa_G=\int d^2 \phi W_G(\phi) \kappa(\phi)
\label{eq:kappaG}
\end{equation}
\begin{equation}
W_G(\phi)=\frac{1}{\pi \theta_G^2}\exp\left(-\frac{\phi^2}{\theta_G^2}\right),
\end{equation}
where the center of the smoothing kernel is set to that of the
cluster. Here $\theta$ is the angular distance from the cluster
center, and $\theta_G$ is the size of the smoothing aperture, which we
choose to be 1 arcmin. Assuming that along the line of sight, there is
only one lens at redshift $z$ (i.e., the cluster to be detected), the
convergence can be calculated, given a distribution of the background
galaxies, as
\begin{equation}
\kappa(\phi)=\frac{4\pi
G}{c^2}\frac{\Sigma(\phi)\chi_z}{(1+z)}
\frac{\int_{z}^{\infty} dz'(dn/dz')(1-\chi_z/\chi_{z'})}{n_{\rm tot}}.
\end{equation}
Here $\Sigma(\phi)$ is the surface density of the cluster, projected
along the line of sight (see~\cite{H&T&Y03} for more details),
$\chi_z$ is the comoving radial distance to redshift $z$ (note that
we assume a flat universe in this paper), $dn/dz$ is the mean
redshift distribution of the surface number density (per steradian)
of background galaxies, and $n_{\rm tot}$ is the mean total surface
density. 

Throughout this paper, we adopt both $dn/dz$ and $n_{\rm tot}$ from
the ground based survey described in ~\cite{S&K04}.  In particular,
those authors assumed $n_{\rm tot}=65~{\rm arcmin^{-2}}$, which may be
optimistic by a factor of $\sim$two~(e.g. \cite{M&B06}, see also
ref.\cite{B&J02} for an extended discussion on statistical galaxy
shape measurements). In this paper, we are primarily interested in a
relative improvement of constraints when two methods are combined,
rather than in the absolute constraints. We therefore stick to the
optimistic values, to facilitate a comparison with earlier work.  We
assume the noise on the measured convergence comes from the
r.m.s. intrinsic ellipticity of the background galaxies, and is given
by ~\cite{LVW00}
\begin{equation}
\sigma^2_{\rm noise}=\frac{\sigma^2_{\epsilon}}{4}\frac{1}{2\pi
\theta_G^2 n_{\rm tot}}.
\end{equation}
Here $\sigma_{\epsilon}$ is the weighted average of the
r.m.s. intrinsic ellipticity per component of the galaxies,
given by
\begin{equation}
\sigma^2_{\epsilon}=\frac{\int_0^{\infty}dz (dn/dz)
\sigma^2_{\epsilon}(z)}{n_{\rm tot}},
\end{equation}
where we follow ~\cite{S&K04}, and take $\sigma_{\epsilon}(z)$ to be
\begin{equation}
\sigma_{\epsilon}(z)=0.3+0.07z.
\label{eq:sigmaez}
\end{equation}

Finally, we set the threshold convergence $\kappa_{\rm th}$ to be 4.5
times the noise. Setting $\kappa_G$ in equation~(\ref{eq:kappaG})
equal to $4.5\sigma_{\rm noise}$ yields an implicit equation for
$M_{200}(z)$.  To be consistent with the halo mass defined for the
Jenkins {\it et al.}  fitting formula for the cluster mass function,
we extend the NFW density profile for a cluster with mass $M_{200}(z)$
outward to a radius so that the mean interior density is 180 times
that of the background matter density at redshift $z$.  The mass
enclosed within this radius is adopted as $M_{\rm min}(z)$ in
equation~(\ref{eqn:Ni}).

In the mock survey defined above (and with the fiducial cosmological
parameters defined below), we find that the limiting halo mass is
$\sim (0.6-4)\times 10^{14}~{\rm M_\odot}$, depending on redshift.
The smallest halos therefore correspond to small groups, containing,
on average, $\sim$(10-60) galaxies above the absolute $r$ magnitude
$M_r\lsim -20$ threshold for the Sloan Digital Sky Survey~(see Figure
3 in ref.\cite{T&W&W06}).  Note that LSST will detect galaxies to a
much greater depth, which may increase the number of detectable member
galaxies (although this is unclear; see, e.g., Figure 3 in
ref.~\cite{AC06} that shows no increase).

The total number of clusters in the survey down to this mass threshold
is $N_{\rm total}=276,794$.  This is somewhat larger than the number
in our previous study~\cite{SW04}. The reason for the increase is that
in the previous paper, we excluded background galaxies at $z>2.5$,
whereas here we include them (and both distributions were normalized
to have the same $n_{\rm tot}$).  As a result, we find here an
increase in the convergence produced by a given foreground galaxy
cluster. Two additional, smaller differences are that here we adopt a
smoothing filter size of 1 arcmin (rather than 0.5 arcmin), and that
we use a slightly higher, redshift--dependent intrinsic ellipticity
(given by eq.(\ref{eq:sigmaez}), rather than a fixed value of
$0.3\sqrt{2}$).

\subsection{Shear--Shear Correlations}
\label{sec:II B}

Here we closely follow ref.~\cite{S&K04}, and consider the
shear--shear correlation function in an LSST--like ground based
survey.  From $z=0$ to $z=3.2$, we divide the background galaxies into
8 equally--spaced redshift bins, and we imagine that the solid angle
covered in the survey is probed by pixels of size of $\Omega_{\rm
pix}$. On average, a given pixel will probe $N_{\rm pix}^b$ galaxies
in redshift bin '$b$', with $N_{\rm pix}^b=n^b\Omega_{\rm pix}$, where
$n^b$ is the mean surface density of the galaxies in this redshift
bin, and is given by

\begin{equation}
n^b=\int_{z_{\rm min}^b}^{z_{\rm max}^b}dz(dn/dz)
\end{equation}
with $z_{\rm min}^b$ and $z_{\rm max}^b$ the edges of this redshift
bin. 

The detected shear from these galaxies depends both on the
lensing shear signal and on their intrinsic ellipticity

\begin{eqnarray}
\gamma_{1,{\rm pix}}^b=\gamma_{1, {\rm lens}}+\gamma_{1, {\rm
int}}=\frac{1}{N_{\rm pix}^b}\sum_{i=1}^{N_{\rm pix}^b}
\left(\frac{\epsilon_{+i,{\rm lens}}}{2}+\frac{\epsilon_{+i,{\rm int}}}{2}\right)\\
\gamma_{2,{\rm pix}}^b=\gamma_{2, {\rm lens}}+\gamma_{2,{\rm
int}}=\frac{1}{N_{\rm pix}^b}\sum_{i=1}^{N_{\rm pix}^b}
\left(\frac{\epsilon_{\times i,{\rm
lens}}}{2}+\frac{\epsilon_{\times i,{\rm int}}}{2}\right)
\end{eqnarray}
where, $\gamma_1$ $\gamma_2$ are the two independent components of the
shear field, $\epsilon_+$ $\epsilon_{\times}$ are the two components
of the galaxy ellipticity.  As before, the intrinsic ellipticity is
the only source of noise we consider in this paper for the shear-shear
correlations.  Furthermore, we assume that there is no correlation
between intrinsic ellipticity of different galaxies.  In this case,
$\gamma_{1,{\rm int}}$, when averaged over large enough realizations
of the intrinsic ellipticity of these $N_{\rm pix}^b$ galaxies, will
have a variance of
\begin{eqnarray}
\langle\gamma_{1,{\rm int}}^2\rangle & = & \frac{1}{4(N_{\rm
pix}^b)^2}\left\langle\sum_i \epsilon_{+i,{\rm
int}}^2\right\rangle\nonumber \\ & = & \frac {\int_{z_{\rm
min}^b}^{z_{\rm max}^b}dz(dn/dz)\langle\epsilon_{+,{\rm
int}}^2(z)\rangle} {4N_{\rm pix}^b n^b}, \label{eqn:shvar}
\end{eqnarray}
A similar expression holds for $\gamma_{2,{\rm int}}$. Following
~\cite{S&K04} again, we assume that the r.m.s. intrinsic ellipticity
varies with redshift as
\begin{equation}
\langle\epsilon_{+,{\rm
int}}^2(z)\rangle=\langle\epsilon_{\times,{\rm
int}}^2(z)\rangle=\sigma^2_{\epsilon}(z),
\end{equation}
where $\sigma_{\epsilon}(z)$ is given by
equation~(\ref{eq:sigmaez}).

We assume further that there is no correlation between the lensing
signal and noise, and expand the shear fields in terms of spherical
harmonics. The covariance matrix $\mathbf{C}^{\gamma \gamma}$ of the
expansion coefficients in the E mode can then be written as the sum
of two matrices $\mathbf{S}^{\gamma \gamma}$(signal) and
$\mathbf{N}^{\gamma \gamma}$(noise), whose elements are
\begin{eqnarray}
S_{\ell m,\ell' m'}^{b,b'}&=&C_\ell^{bb'}\delta_{\ell \ell'}\delta_{mm'} \\
N_{\ell m,\ell' m'}^{b,b'}&=&N^b\delta_{bb'}\delta_{\ell
\ell'}\delta_{mm'}
\end{eqnarray}
where $b, b'$ label redshift bins, and $\ell, m, \ell' ,m'$ label
spherical harmonic modes. The shear angular power spectrum is then
given by
\begin{equation}
C_\ell^{bb'}=\frac{\pi^2 \ell}{2}\int_0^{\infty}dz
\frac{d\chi_z}{dz} \frac{W^b(z)W^{b'}(z)}{\chi_z
^3}\Delta^2_{\Phi}(k,z) \label{eqn:cl}
\end{equation}
where $k=\ell/\chi_z$, and $\Delta^2_{\Phi}(k,z)$ is the variance of
the gravitational potential fluctuations per $\ln k$ interval. The
window function is given by
\begin{equation}
W^b(z)=\frac{2\chi_z}{n^b} \int_{z_{\rm min}^b}^{z_{\rm max}^b}
dz'\frac{dn}{dz'} \left(1-\frac{\chi_z}{\chi_{z'}}\right)
\Theta(z'-z) \label{eq:zwindow}
\end{equation}
where $\Theta$ is the step function. The diagonal elements of the
noise matrix $N$ are $\ell, m$ independent (although they depend on
the redshift bin), and given by
\begin{equation}
N^b=\langle\gamma_{{\rm int}}^2\rangle_b\Omega_{\rm pix},
\end{equation}
where $\langle\gamma_{{\rm int}}^2 \rangle_b$ refers either to
$\langle\gamma_{1,{\rm int}}^2\rangle$ or to $\langle\gamma_{2,{\rm
int}}^2\rangle$, as given by equation~(\ref{eqn:shvar}).

In the late universe, $\Delta^2_{\Phi}(k,z)$ is simply related to
$\Delta^2(k,z)$, the variance of matter density fluctuations per
$\ln k$ interval, by
\begin{equation}
\Delta^2_{\Phi}(k,z)= \left(\frac{3\Omega_m}{2a}\right)^2
\left(\frac{H_0}{ck}\right)^4 \Delta^2(k,z)
\end{equation}
where $\Omega_m$ is the present--day matter density parameter, $H_0$
is the present--day Hubble constant, $a$ is the scale factor
normalized to unity today, and $c$ is the speed of light. 

This equation neglects fluctuations in dark energy. For a
scalar-modeled dark energy, this is assured only for wavelengths
shorter than the Compton wavelength of the scalar field, or $k\gg
k_Q$, where $k_Q$ is the Compton wavenumber. Ma {\it et al.}
~\cite{CPM99} gave a fitting formula for $k_Q(z)$ when the scalar
field equation of state parameter $w$ is a constant and the universe
is spatially flat,
\begin{equation}
k_Q=\frac{3H(a)}{c}\sqrt{(1-w)
\left[2+2w-\frac{w\Omega_m}{\Omega_m+(1-\Omega_m)a^{-3w}}\right]}.
\end{equation}
Since $k=\ell/\chi$, we require $\ell\gg k_Q \chi$ for
self--consistency. In a cosmological model close to our fiducial model
(defined in the next section) but with $w\rightarrow -1$, the maximum
of $k_Q \chi$ is $\approx 30$.  We follow ~\cite{S&K04} and impose an
upper limit on the angular scales utilized in our study, given by
$\ell>40$.  Note, however, that dark energy fluctuations are unlikely
to be actually detectable on large scales, and our results are
insensitive to this lower limit on $\ell$ (see discussion below).

\subsection{Error Estimates}

We assume the spatial curvature of the universe is zero, and adopt a
7--dimensional cosmological parameter set $\{\Omega_{\rm
DE},\Omega_mh^2,\sigma_8, w_0,w_a,\Omega_b h^2,n_s\}$, where
$\Omega_{\rm DE},\Omega_m,\Omega_b$ are present--day energy density
parameters of dark energy, total matter (cold dark matter+baryon), and
baryons, respectively, $h$ is the Hubble constant in units of
$100~{\rm km~s^{-1}~Mpc^{-1}}$, $n_s$ is the index for the primordial
matter power spectrum, and $\sigma_8$ is the amplitude of the linear
matter density fluctuations today smoothed on a scale of
$8h^{-1}$Mpc. We consider a time--varying equation of state parameter,
given by
\begin{equation}
w(a)=w_0+w_a(1-a)
\end{equation}
In our fiducial model, we choose the following values of the 7
parameters: $\{0.73,0.14,0.9,-1,0,0.024,1\}$, which are consistent
with the current ``concordance model''~\cite{WMAP3}.  

To estimate uncertainties on the cosmological parameters obtained by a
specific probe, we use the Fisher matrix formalism. The Fisher matrix
is defined as,
\begin{equation}
F_{\alpha \beta}= -\left\langle\frac{\partial^2 \ln L}{\partial
p_{\alpha}\partial p_{\beta}}\right\rangle \label{eq:Fishermatrix}
\end{equation}
where $p_{\alpha}, p_{\beta}$ represent the 7 model parameters we
want to constrain, $L$ is the likelihood function, the derivatives
are evaluated at the true parameter set (which in our case is the
adopted fiducial parameter set), and the average is taken over many
realizations of the data set. The uncertainty on the parameter
$p_{\alpha}$ after marginalized over all other parameters is
obtained as $\sqrt{(F^{-1})_{\alpha\alpha}}$, which gives a lower
limit on the accuracy of $p_\alpha$ for any unbiased estimator of
the parameters~\cite{T&T&H97}, and in the absence of systematic
errors.

For number counts, we assume that the counts in each of the 26
redshift bins are independent Poisson random variables, with the
mean values given by equation~(\ref{eqn:Ni}).  The Fisher matrix in
this case is constructed as ~\cite{H&H&M01},
\begin{equation}
F_{\alpha\beta}^{\rm counts}=\sum_{i=1}^{26}\frac{\partial\bar{
N_i}}{\partial p_{\alpha}}\frac{\partial \bar{N_i}}{\partial
p_{\beta}}\frac{1}{\bar{N_i}}. \label{eq:Fishercounts}
\end{equation}

For the shear--shear correlations, we assume that the spherical
harmonic expansion coefficients are Gaussian random variables whose
mean are zero, and whose covariance matrix is $\mathbf{C}^{\gamma
\gamma}$. The Fisher matrix is then constructed as~\cite{K&C&E&H01}
\begin{equation}
F_{\alpha\beta}^{\gamma \gamma}=\frac{1}{2}f_{\rm
sky}\sum_{\ell,b_1,b_2,\atop
b_3,b_4}(2\ell+1)C_{\ell,\alpha}^{b_1b_2}W_\ell^{b_2b_3}C_{\ell,\beta}^{b_3b_4}W_\ell^{b_4b_1}
\label{eq:Fishershear}
\end{equation}
where $f_{\rm sky}$ is the fraction of the sky coved by the survey
(in our case, $f_{\rm sky}=0.44$), $[X]_{,\alpha}$ denotes the
derivative of $[X]$ with respect to $p_{\alpha}$, $C_\ell^{bb'}$ is
calculated from equation~(\ref{eqn:cl}), and the $W_\ell^{bb'}$ are
the elements of the inverse of the covariance matrix,
\begin{eqnarray}
\mathbf{W}&=&(\mathbf{C}^{\gamma \gamma})^{-1} \\
W_{\ell
m,\ell'm'}^{b,b'}&=&W_\ell^{bb'}\delta_{\ell\ell'}\delta_{mm'}.
\end{eqnarray}

Non--linear gravitational clustering in the late universe will induce
non--Gaussian signatures in matter fluctuation field.  Although we
incorporate these non--linear effects on the matter power spectrum,
these non--linear effects will also render the likelihood function
non--Gaussian.  To avoid this complication, we neglect all modes with
$\ell>1000$ ~\cite{S&Z&H99}.

To compute the linear matter power spectrum, we use KINKFAST
~\cite{PSC04}, a version of CMBFAST ~\cite{S&Z96} modified to
accommodate a time--varying equation of state parameter $w$, to
calculate the transfer function at $z=0$, and we obtain the linear
growth function by integrating the differential equations given in
the Appendix of ref.~\cite{W&S98}. For the shear--shear
correlations, the non--linear matter power spectrum is constructed
following Smith {\it et al.}~\cite{RES03}. The derivatives in the
Fisher matrices are calculated by two-sided numerical
approximations.  For both $F^{\rm counts}$ and $F^{\gamma\gamma}$,
we chose a step--size of $\Delta w_a=\pm 0.01$, and $\pm1\%$ of the
fiducial value for the other 6 parameters.  We have verified
directly that these step--sizes are small enough for the Fisher
matrix entries to have converged.

\subsection{Covariance Between Number Counts and Shear}
\label{sec:covariance}

Our treatments of the number counts (only shot noise is considered)
and its combination with shear-shear correlations (covariance between
these two are neglected) are quite simplified. To yield more accurate
predictions, sample variance errors for the number counts, and the
covariance between number counts and shear-shear correlations should
be taken into account.

The effect of sample variance on the constraining power of number
counts has been considered in detail in previous
work~\cite{H&K03,L&H04}.  In particular, ref.~\cite{H&K03} finds (see
their Figure 2) that the degradation on dark energy constraints, when
the sample variance error is added to the Poisson error, depends
mostly on the mass threshold. At the lower end of the range of our
fiducial mass thresholds, $\sim (0.6-4)\times 10^{14}~{\rm M_\odot}$,
the degradation is a factor of $\sim 2$, while at the upper end, there
is negligible degradation (see their Figure 7).  However, these
degradations are overestimates, because ref.~\cite{H&K03} excluded the
``signal'' that arises from the cosmology dependence of the sample
variance.  When this information is included, sample variance errors
should cause a smaller degradation in the number count constraints
(and possibly even improvement, if the survey is sub--divided into
many angular cells, as in \cite{L&H04}).

For our purposes, the more important question is whether the
covariance between number counts and shear--shear correlations is
significant.  The potential concern is that the shear--shear
correlations probe the same realization of the density field as the
cluster counts, and therefore simply adding the constraints from the
two observables may overestimate their combined constraining
power. Indeed, in the hypothetical limit that the mean cluster
abundance in a particular direction and redshift bin is fully
predictable, given measurements of the shear correlations, the cluster
counts should not yield any new information on dark energy. However,
we show here that the covariance is very small, and the two
observables can safely be regarded as independent.  In this section,
we summarize the results of the covariance calculation; the interested
reader is encouraged to consult the Appendix for details.

In the general case, the Fisher matrix, defined in
equation~(\ref{eq:Fishermatrix}), involves the expectation value of
the derivatives of the joint likelihood function $L=L({\mathbf
x},{\mathbf p})$ where ${\mathbf x}$ is a vector of the observables,
and ${\mathbf p}$ is a vector of the model parameters.  In our case,
${\mathbf x}$ contains the number counts in 26 redshift bins, $\{N_i;
1 \leq i \leq 26\}$, and the spherical expansion coefficients in 8
different redshift bins $\{a_{\ell m}^b; 1\leq B\leq 8, 41\leq \ell
\leq 1000, -(\ell+1)\leq m_\ell \leq (\ell+1)\}$ (note that we have
8,002,560 + 26 = 8,002,586 observables).  Here both $N_i$ and $a_{\ell
m}^b$ are random variables, and in the discussion below, the
probability distribution of both are taken to be determined by
large--scale density fluctuations alone. In practice, the measurement
of either quantity represents a discrete sampling of a continuous
random field, and will therefore have an additional sampling error.
In particular, equation~(\ref{eq:Fishercounts}) assumes that Poisson
errors dominate the sample variance errors, whereas
equation~(\ref{eq:Fishershear}) incorporates the additional stochastic
noise from the distribution if intrinsic shapes.  However, such
sampling errors should be uncorrelated and will be ignored below.
Note that excluding truly uncorrelated errors is conservative, since
they would reduce the cross--correlation coefficient defined below
(eq.\ref{eq:cross}).

Under the assumption that the full joint likelihood function is
Gaussian, the Fisher matrix depends only on the mean
$\mathbf{\bar{x}}$ and the covariance matrix $\mathbf{C}=
\langle(\mathbf{x}-\mathbf{\bar{x}})
(\mathbf{x}-\mathbf{\bar{x}})^T\rangle$, and on the derivatives of
these quantities with respect to the model parameters
$\mathbf{p}$~\cite{T&T&H97} (note that in general, $\mathbf{\bar{x}}$
and $\mathbf{C}$ both depend on $\mathbf{p}$).  In our case, the full
covariance matrix ${\bf C}$ contains the terms $\langle(N_i-\bar{N_i})
(N_j-\bar{N_j})\rangle$ and $\langle a_{\ell m}^b a_{\ell^\prime
m^\prime}^{b^\prime*}\rangle$, which describe the sample variance in
the number counts and in the shear field, respectively.
\footnote{Note that eq.~(\ref{eq:Fishershear}) depends on $C_\ell$ and
its derivatives, rather than $a_{\ell m}$.  This dependence arises
from taking the expectation value $C_\ell\equiv [1/(2\ell+1)]\langle
\Sigma_{m=-\ell}^{+\ell} |a_{\ell m}|^2\rangle$.}  The cross--terms,
$\langle(N_i-\bar{N_i})a_{\ell m}^b\rangle$, describe the covariance
between number counts and the shear field.  Here $\bar{N_i}$ is the
mean number of clusters given in equation~(\ref{eqn:Ni}), while we
have $\overline{a_{\ell m}^b}=0$, and the averages are taken over many
realization, or a large volume.

In the Appendix, we calculate the cross--terms explicitly, and show
that they are given by a simple expression,
\begin{multline}
\langle(N_i-\bar{N_i})a_{\ell m}^b\rangle =
\delta_{m0}\frac{3\pi^2}{2}\frac{\Omega_m H_0^2}{c^2}
\frac{\chi_{z_i}(1+z_i)}{\ell^3}\times \\
\tilde{\Theta_\ell}N_i
b(z_i)W^b(z_i)\Delta^2(k=\frac{\ell}{\chi_{z_i}},z_i)\label{eqn:cov-ns}.
\end{multline}
Here $\tilde{\Theta_\ell}$ is the spherical harmonic transform of an
azimuthally symmetric angular window function, $W^b$ is the lensing
window function (given in equation~\ref{eq:zwindow}), and $b(z)$ is
the mean bias factor of the cluster counts (averaged over clusters
above the detection threshold).  The above result assumes that the
cluster number counts trace the matter density field with the linear
bias factor $b$, and we have also used the Limber approximation.  The
latter assumption should be justified for the angular modes we use
($\ell>40$).  Note that the cross--term vanishes for clusters behind
the source galaxies (since $W^b(z_i)=0$ for $z_i>z_{\rm max}^b$), and
also for $m\neq 0$ (since our survey window is azimuthally symmetric).

\begin{figure}[t!]
 \resizebox{90mm}{!}{\includegraphics{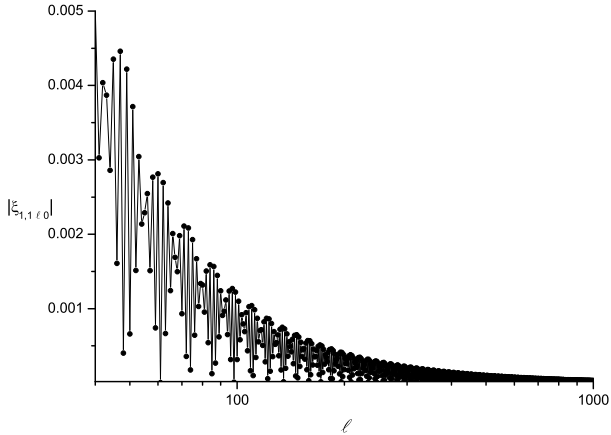}}
 \caption{\label{fig:4a} Absolute values of the cross--correlation
coefficients (defined in eq.\ref{eq:cross} in the text) between the
number of galaxy clusters in redshift range of $(0.1<z<0.15)$ and the
spherical harmonic expansion coefficients (at $m$=0) of the shear map
from source galaxies in redshift range of $(0<z<0.4)$. The figure
shows that the cross--correlation coefficients are of order $10^{-3}$
or smaller for $\ell>40$.}
\end{figure}

We define the cross--correlation coefficient between the $N_i$ and the
$a_{\ell m}^b$ as
\begin{equation}
\xi_{i,b \ell m} \equiv \frac{\langle(N_i-\bar{N_i}) a_{\ell
m}^b\rangle}{\sqrt{\langle (N_i-\bar{N_i})^2 \rangle \langle |a_{\ell
m}^b|^2 \rangle}}. \label{eq:cross}
\end{equation}
As an example, here we calculate the cross--correlation coefficient
for the number counts and the shear field in their lowest respective
redshift bins, $|\xi_{1,1 \ell 0}|$.  The results are shown, as a
function of $\ell$, in Figure~\ref{fig:4a}.  The figure shows that
the $|\xi_{1,1 \ell 0}|$ are small -- of order $10^{-3}$ or less --
for $\ell>40$.  For a different pair of redshifts, $i$ and $B$, we
expect this order of magnitude would not change significantly, since
the related quantities vary slowly with redshift.

\begin{figure}[t!]
 \resizebox{90mm}{!}{\includegraphics{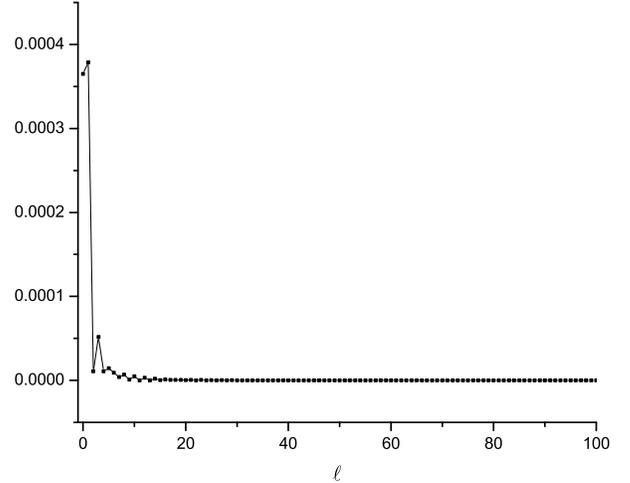}}
 \caption{\label{fig:4b} Contributions from different $\ell$--modes to
   the sample variance for the number of galaxy clusters in the first
   redshift bin ($0.1 < z < 0.15$), containing, on average, 7515
   clusters.  The figure shows that the sample variance is dominated
   by the largest angular scales. This explains the lack of
   cross--correlation between the counts and the shear field, when the
   latter is restricted to smaller angular scales ($\ell>40$).}
\end{figure}

The fact that the $\xi_{i,b \ell m}$ are small can be explained by the
following reasoning. For a given survey window, the variance in the
cluster counts alone is dominated by the largest angular modes, due to
cancellations among smaller--scale fluctuations along the direction
transverse to the line of sight.  In Figure~\ref{fig:4b}, we
explicitly show the contributions to the sample variance of the
cluster counts in the first redshift bin for LSST (note that we do not
use the Limber approximation for this calculation).  Because of the
large angular size of the window, the figure clearly shows that the
variance is small for modes with $\ell\gsim 10$.  Fluctuations in the
underlying (isotropic) matter density field on different scales $\ell$
are uncorrelated, and we have limited the range of angular modes of
the shear maps to $\ell>40$.  Since these relatively smaller--scale
modes contribute little to the fluctuations in the number counts, we
indeed expect the $\xi_{i,b \ell m}$ to be small.

According to these results, the probability of drawing a set of $N_i$
and $a_{\ell,m}^b$ is given by a multi--variate Gaussian, with the
total covariance matrix that consists of four blocks,
\begin{equation}
\mathbf{C_{\rm tot}}=\left( \begin{matrix} \mathbf{S^{\rm counts}}& \mathbf{C^{\rm cross}}\\
\mathbf{(C^{\rm cross})^T}& \mathbf{S^{\gamma\gamma}}
\end{matrix}\right),
\end{equation}
where $\mathbf{S^{\rm counts}}$ and $\mathbf{S^{\gamma\gamma}}$ are
the sample variance matrices for counts and shear alone. When
$|\xi_{i,b \ell m}|\ll 1$, this matrix can be well approximated (for
example, for the purpose of taking its inverse) by
\begin{equation}
\mathbf{C_{\rm tot}}=\left( \begin{matrix} \mathbf{S^{\rm counts}}& \mathbf{0}\\
\mathbf{0}& \mathbf{S^{\gamma\gamma}} \end{matrix}\right).
\end{equation}
A full treatment would incorporate the Poisson errors for the counts,
and the shape--errors for the shear, however, these effects are
relatively small, and could only decrease the value of $\xi$.  The
Fisher matrix also requires taking the derivatives with respect to the
cosmological parameters.  As long as these derivatives do not have a
strong scale-dependence, and given the smallness of $\xi$, we expect
our basic conclusion to carry over to the Fisher matrix.  Therefore,
the cluster number counts and shear-shear correlations can be treated
as independent probes, even if they probe the same area of the sky.
For the results below, we therefore simply add the two Fisher
matrices, when we combine the two observables.

\section{Results}
\label{sec:III}

The marginalized errors on the seven cosmological parameters from the
number counts and the shear-shear correlations are listed in the
$2^{\rm nd}$ and $3^{\rm rd}$ columns of Table~\ref{tab:main} for our
fiducial LSST--like survey.  Note that the results scale simply as
$\Delta\Omega^{-1/2}$ for a survey with a different solid angle
coverage.  The $4^{\rm th}$ column shows the result from combining the
two observable, assuming that they are independent, so that the two
Fisher matrices can simply be added (see the discussion of the
covariance between these two probes in \S~\ref{sec:covariance} above).
In the limit that the two observable have the same degeneracies
between parameters, their combination would be equivalent to simply
adding the marginalized errors in quadrature. In the $5^{\rm th}$
column of Table~\ref{tab:main}, we show a ``complementarity''
parameter, $\eta$, which quantifies the effect of degeneracy--breaking
that occurs when the two methods are combined.  The parameter $\eta$
is defined as the improvement of the constraints beyond adding the two
results in quadrature,
\begin{equation}
\eta=(\Delta p_{\alpha}^{{\rm counts}+\gamma\gamma})^2
\left[\frac{1}{(\Delta
p_{\alpha}^{\rm counts})^2}+\frac{1}{(\Delta
p_{\alpha}^{\gamma\gamma})^2}\right].
\end{equation}
With this definition, $\eta=1$ corresponds to no degeneracy breaking,
and lower values of $\eta$ indicate larger benefits from the
combination. The $6^{\rm th}$ and $7^{\rm th}$ columns are the same as
the $3^{\rm rd}$ and $4^{\rm th}$, except that, as an academic
exercise, for the shear-shear correlations, we use the linear matter
power spectrum instead of the nonlinear one (see the next section for
a detailed discussion).

\begin{table}
  \caption{\label{tab:main} {\it This table contains our main
    results.}  Marginalized errors are shown on cosmological
    parameters from cluster counts, shear--shear correlations, and
    their combination. The parameter $\eta$, shown in the $5^{\rm th}$
    column, measures the synergy between the two observable, with
    $\eta=1$ indicating no synergy, and lower values indicating
    significant degeneracy--breaking (see text for definition).  In
    the $6^{\rm th}$ and $7^{\rm th}$ columns, we use the linear power
    spectrum for the shear (indicated here, and in the other tables
    below, by the superscript ``l'').  Priors from WMAP,
    $\Delta\Omega_b h^2=0.0010, \Delta n_s=0.040$, are included for
    the $2^{\rm nd}, 4^{\rm th}$, and $7^{\rm th}$ columns here, and
    in columns involving number counts in all other tables below
    (except Table~\ref{tab:pshear}).  }
\begin{tabular}{|c|c|c|c|c|c|c|}
\hline
 &counts&$\gamma \gamma$&{{\rm counts}+$\gamma \gamma$}&$\eta$&$\gamma \gamma^l$ &{{\rm counts}+$\gamma \gamma^l$}\\
\hline
$\Delta \Omega_{\rm DE}$ &0.064   &0.0084   &0.0026    &0.097   &0.013     &0.0027\\
$\Delta \Omega_m h^2$     &0.20    &0.049    &0.0061    &0.017   &0.034     &0.0050\\
$\Delta \sigma_8$         &0.029   &0.012    &0.0031    &0.083   &0.021     &0.0031\\
$\Delta w_0$              &0.080   &0.078    &0.033     &0.34    &0.12      &0.031\\
$\Delta w_a$              &1.24    &0.28     &0.11      &0.18    &0.42      &0.099\\
$\Delta \Omega_b h^2$     &0.0010  &0.014    &0.00099   &0.99    &0.010     &0.00099\\
$\Delta n_s$              &0.040   &0.050    &0.016     &0.26    &0.027     &0.011\\
\hline
\end{tabular}
\end{table}

The first conclusion to draw from Table~\ref{tab:main} is that the
shear--shear correlations give tighter constraints on all cosmological
parameters than cluster counts alone, especially on $\Omega_\Lambda,
\Omega_m h^2,\sigma_8$ and $w_a$. This may not be surprising, given
that number counts effectively measure only 26 numbers, while the
shear power spectrum is effectively a measurement of many more
parameters. On the other hand, cluster counts deliver a constraint
comparable to that from the shear for $w_0$.  We also note that
constraints from cluster counts alone on $w_a$ are weak (as noted by
~\cite{SW04}, this can be significantly improved by adding the cluster
power spectrum and CMB anisotropy as observable).\footnote{We note for
reference that our shear correlation constraints are consistent with a
slightly updated version of the results in ref.~\cite{S&K04}.  Our
number--count--alone constraints on $\Omega_{\rm DE}$ and $w_a$, on
the other hand, are significantly weaker than in ref.~\cite{SW04}. We
have found that the discrepancy is due to inaccurate interpolation in
ref.~\cite{SW04} to obtain the mass limit.  We have also found,
however, that the inaccuracies do not significantly alter the joint
constraints when the number counts are combined with other
observables.}

\begin{figure}[t!]
 \resizebox{90mm}{!}{\includegraphics{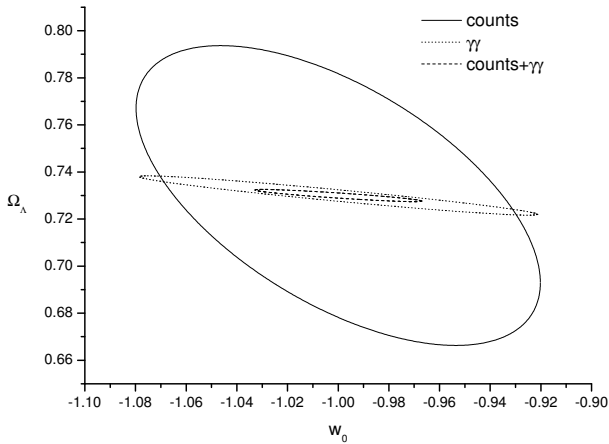}}
 \caption{\label{fig:1} Marginalized constraints in the ($\Omega_{\rm
     DE}, w_0$) plane for an LSST--like survey from the
   shear--selected cluster counts, the shear--shear correlations, and
   the combination of these two observables. Note that the cluster
   counts alone deliver weaker constraints, but still improve the
   $w_0$ errors from the shear--shear correlations by a factor of
   $\sim$two, as a result of breaking degeneracies.}
\end{figure}

\begin{figure}[t!]
 \resizebox{90mm}{!}{\includegraphics{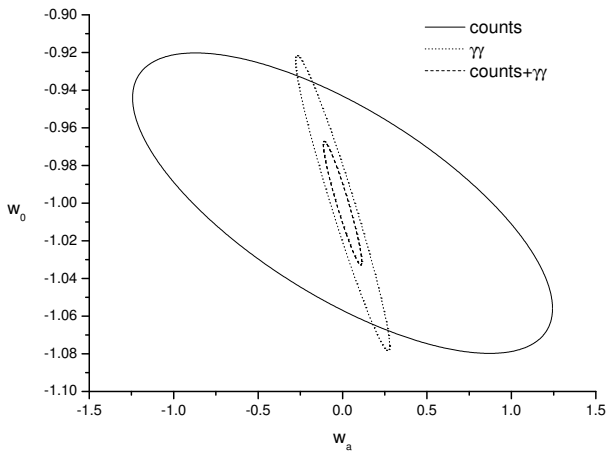}}
 \caption{\label{fig:2} Marginalized constraints in the ($w_0, w_a$)
   plane for an LSST--like survey from the shear--selected cluster
   counts, the shear--shear correlations, and the combination of these
   two observables.}
\end{figure}

More importantly, Table~\ref{tab:main} shows that when the two methods
are combined, the constraints on the cosmological parameters improve
significantly. For most of the parameters, $\eta$ is small, indicating
significant complementarity.  In particular, focusing on the three
dark--energy parameters, the combination tightens the constraints by a
factor of $3-10$ more than simply adding the marginalized errors in
quadrature.  The combined constraint on $\Omega_{\rm DE}$ is $\sim$ 25
times better than that from counts and $\sim$3 times better than that
from $\gamma\gamma$; the combined constraint on $w_0$ is $\sim$ 2
times better than either from counts or $\gamma\gamma$; and the
combined constraint on $w_a$ is $\sim$ 11 times better than from
counts and $\sim$ 3 times better than from $\gamma\gamma$. These
results are also shown graphically in Figures~\ref{fig:1} and
~\ref{fig:2}.

For reference, we follow the recommendation of the Dark Energy Task
Force (DETF)~\cite{DETF}, and compute the ``pivot point''--$a_p$,
i.e. the value of the scale factor where $w(a)$ is best constrained.
In Table ~\ref{tab:pivot}, we list the values of the scale factor
and redshift of this pivot point, as well as the errors on $w(a_p)$
and the ``figure of merit'' defined by the DETF, $(\Delta w_p \Delta
w_a)^{-1}$.  The figure of merit we find for the individual
observable is in-between the ``optimistic'' and ``pessimistic''
predictions by the DETF.  The combined figure of merit, however, is
significantly better than the most optimistic figure of merit in the
DETF for a ``stage II'' WL shear experiment alone.
\begin{table}
\caption{\label{tab:pivot} {\it This table recast our results in terms
of a pivot point}. The pivot point is defined as the scale factor at
which the equation of state $w_p\equiv w(a_p)$ is best constrained.
The last row shows the figure--of--merit proposed by the
DETF~\cite{DETF}.}
\begin{tabular}{|c|c|c|c|c|}
\hline
 &counts&$\gamma \gamma$&{{\rm counts}+$\gamma \gamma$}\\
\hline
$z_p$  &0.047& 0.37 &0.37\\
$1-a_p$   &0.045   &0.27    &0.27\\
$\Delta w_p$ &0.057   &0.020   &0.010  \\
$\Delta w_a$   &1.24    &0.28     &0.11\\
$(\Delta w_p \Delta w_a)^{-1}$    &14.1   &178.6    &909.1  \\
\hline
\end{tabular}
\end{table}

\section{Discussion}
\label{sec:IV} 

In this section, we discuss several aspects of our basic simple
results presented in the previous section. In particular, we explain
the reasons for the synergy between the two observable, and discuss
various possible caveats that could modify our conclusions.

\subsection{Degeneracy Breaking by the Two Observables}

Table~\ref{tab:unmar} lists the uncertainties on the 7 cosmological
parameters when fixing all the other 6 parameters at their fiducial
value. By comparing Table~\ref{tab:unmar} with Table~\ref{tab:main},
we see that, using either observable alone, degeneracies among the 7
parameters lead to great degradation on the constraints. Clearly,
combining these two observables can decrease the effect of this
degradation by breaking degeneracies. To understand this better, we
find the worst-constrained directions in the 7D parameter space for
both of the observables. We do this by diagonalizing the Fisher
matrix, and finding the eigenvector that corresponds to the smallest
eigenvalue, i.e. whose direction is the one along which the probe is
least sensitive at and hence constrained worst. For the number
counts, we find that the ($\Omega_{\rm
DE},\Omega_mh^2,\sigma_8,w_0,w_a,\Omega_b h^2,n_s$)--components of
the unit vector pointing in this direction are (0.050, 0.15, 0.023,
-0.044, 0.99, 0, -0.00016), with the corresponding eigenvalue of
0.63. This implies that constraints on ($\Omega_{\rm DE}, w_0,w_a$)
can not be better than (0.063, 0.056, 1.24). For shear--shear
correlations, the direction of the worst degeneracy is (0.026,
0.084, 0.038, -0.26, 0.95, 0.022, -0.10), with a corresponding
eigenvalue of 12. Again, this implies that constraints on
($\Omega_{\rm DE}, w_0,w_a$) can not be better than (0.0077, 0.076,
0.28).  In both cases, we find that the $2^{\rm nd}$ worst
eigenvalue is much greater that the worst -- by a factor of $\sim
5^2$ for the shear--shear correlations, and by a factor of $17^2$
for the number counts. Apparently the 7--dimensional error ellipsoid
is very narrow, with a large extension in one direction that nearly
(but not exactly) coincides with the $w_a$ axis.

Given the very severe degeneracies, and the fact that they do not
point in the same direction for the two observables, it is not
surprising that the combination of the two observables leads to a
tightening of the constraints beyond adding the marginalized errors in
quadrature.

\begin{table}
  \caption{\label{tab:unmar} {\it This table shows the effect of
    marginalization.} The parameter errors are shown from the two
    observables, as in Table~\ref{tab:main}, but before
    marginalization (i.e. assuming the other 6 parameters are fixed).}
\begin{tabular}{|c|c|c|}
\hline
 &counts&$\gamma \gamma$\\
\hline
$\Delta \Omega_{\rm DE}$ &0.00036      &0.00041  \\
$\Delta \Omega_m h^2$     &0.0013       &0.0015   \\
$\Delta \sigma_8$         &0.00042      &0.00064   \\
$\Delta w_0$              &0.0054       &0.0060   \\
$\Delta w_a$              &0.029        &0.024    \\
$\Delta \Omega_b h^2$     &0.00050      &0.00060   \\
$\Delta n_s$              &0.0034       &0.0044   \\
\hline
\end{tabular}
\end{table}

\begin{table}
\caption{\label{tab:pshear} {\it This table shows the effect of the
WMAP priors.} Marginalized errors are shown from cluster counts, and
the combination of the counts and the shear--shear correlations, as in
the $2^{\rm nd}$ and $3^{\rm rd}$ columns of Table~\ref{tab:main},
except that we exclude the WMAP priors.}
\begin{tabular}{|c|c|c|}
\hline
 &counts$^{\rm no priors}$&(counts+$\gamma \gamma)^{\rm no priors}$\\
\hline
$\Delta \Omega_{\rm DE}$     &0.15      &0.0027\\
$\Delta \Omega_m h^2$        &3.92      &0.045\\
$\Delta \sigma_8$            &0.067     &0.0034\\
$\Delta w_0$                 &0.20      &0.037\\
$\Delta w_a$                 &1.40      &0.14\\
$\Delta \Omega_b h^2$        &1.46      &0.013\\
$\Delta n_s$                 &5.04      &0.044\\
\hline
\end{tabular}
\end{table}

\begin{table}
  \caption{\label{tab:3000} {\it This table shows the effect of
including shear measurements on small angular scales.}  Marginalized
errors are shown on the cosmological parameters, as in
Table~\ref{tab:main}.  The difference from Table~\ref{tab:main} is
that we have used additional small--scales for the shear--shear
correlations, by increasing the cutoff from $\ell_{\rm max}=1000$ to
$\ell_{\rm max}=3000$.}
\begin{tabular}{|c|c|c|c|c|c|c|}
\hline
 &counts&$\gamma \gamma$&{{\rm counts}+$\gamma \gamma$}&$\eta$&$\gamma \gamma^l$ &{{\rm counts}+$\gamma \gamma^l$}\\
\hline
$\Delta \Omega_{\rm DE}$ &0.064   &0.0046   &0.0023    &0.25     &0.012    &0.0025\\
$\Delta \Omega_m h^2$     &0.20    &0.034    &0.0050    &0.023    &0.021    &0.0039\\
$\Delta \sigma_8$         &0.029   &0.0060   &0.0027    &0.21     &0.019    &0.0029\\
$\Delta w_0$              &0.080   &0.046    &0.025     &0.38     &0.11     &0.028\\
$\Delta w_a$              &1.24    &0.15     &0.079     &0.27     &0.39     &0.084\\
$\Delta \Omega_b h^2$     &0.0010  &0.010    &0.00099   &1.0      &0.0072   &0.00099\\
$\Delta n_s$              &0.040   &0.026    &0.0097    &0.19     &0.011    &0.0054\\
\hline
\end{tabular}
\end{table}

Finally, since we have added WMAP priors for the number counts
constraints, it is useful to ask how important these priors were for
the combined errors.  In Table~\ref{tab:pshear}, we show the
constraints from the cluster counts, and the combination of the counts
and the shear--shear correlations, as in the $2^{\rm nd}$ and $3^{\rm
rd}$ columns of Table~\ref{tab:main}, except that we exclude the WMAP
priors.  As Table~\ref{tab:pshear} shows, the clusters counts are
insensitive to $\Omega_b h^2$ and $n_s$, and degeneracies with these
parameters also degrade the constraints on other parameters.  However,
the shear--shear correlation provides a sufficiently accurate
measurement of $\Omega_b h^2$ and $n_s$, and the WMAP priors are not
important for the combined constraints.

\subsection{Information from Small--Scale Shear--Shear Correlations and Non--Linearities}

In the above results, we have imposed a small--scale
cutoff, $\ell_{\rm max}=1000$ for the shear-shear correlations.  It is
possible, however, at least in principle, to obtain accurate
non--linear shear power spectra, using numerical simulations on a
large grid of cosmological parameters. It is therefore interesting to
ask whether including higher $\ell$ modes could improve the final
results significantly.  To answer this question, we repeated all the
calculations in Table~\ref{tab:main}, but this time with $\ell_{\rm
max}=3000$.  The results are listed in Table \ref{tab:3000}.

First, we notice from either Table \ref{tab:main} or Table
\ref{tab:3000}, that constraints from the shear-shear correlations
alone are better when the non--linear power spectrum is used than
those from the linear version.  This suggests that there is extra
information in the shear--shear correlations that comes from the
non--linear effects.  Furthermore, comparing Table \ref{tab:main} to
Table \ref{tab:3000}, we see that the improvement on dark energy
constraints from these non--linearities is only about $50\%$ for
$\ell_{\rm max}=1000$, but increases to a factor of $2-3$ for
$\ell_{\rm max}=3000$.  However, Tables \ref{tab:main} and
\ref{tab:3000} both show that once the cluster count information is
added, the nonlinear effects on the shear--shear power spectrum become
essentially irrelevant, even for $\ell_{\rm max}=3000$.

This indicates that while non--linear effects change the shear--shear
power spectrum, the information content of these changes is
sub--dominant compared with that probed by number counts, at least for
the non--linearities contained in modes with $\ell\leq3000$ (note that
clusters are strongly nonlinear objects).  Given these results, we are
satisfied to neglect the higher $\ell$ modes and stick to our original
choice of $\ell\leq 1000$. The fact that higher $\ell$ modes
($3000\geq \ell>1000$) help little ($\sim 10\%$) on the linear
shear--shear correlations, but help more ($\sim$ 50\%) with the
nonlinear shear-shear correlations, together with the fact that adding
higher $\ell$ modes help little (30\% at most) in either case when the
number counts are included, is again an indication that the nonlinear
evolution information in these modes is sub--dominant compared to the
information contained in number counts.

\subsection{Large Scale Shear--Shear Correlations and Dark Energy Clustering}

The previous sub--section showed that while most of the information
in the non--linear shear--shear correlations is on small scales, the
information contained in the linear shear--shear correlations is
coming mostly from larger scales $\ell < 1000$.  For completeness,
we here ask whether the largest scales ($\ell \sim 40$) actually
dominate shear--shear information. We show, in Table~\ref{tab:100},
the constraints, recalculated as in Table~\ref{tab:main}, except we
have neglected large--scales for the shear--shear correlations, by
increasing the cutoff from $\ell_{\rm min}=41$ to $\ell_{\rm
min}=100$.  As the comparison of the two tables show, the
constraints do not degrade significantly, implying that most of the
information is at smaller angular scales ($\ell > 100$).
\begin{table}
  \caption{\label{tab:100} {\it This table shows the effect of
   excluding shear measurements on the largest scales.} Marginalized
   errors on the cosmological parameters, as in
   Table~\ref{tab:main}. The difference from Table~\ref{tab:main} is
   that we have neglected large--scales for the shear--shear
   correlations, by increasing the cutoff from $\ell_{\rm min}=41$ to
   $\ell_{\rm min}=100$.  As the comparison of the two tables show,
   the constraints do not degrade significantly, implying that most of
   the information is at relatively small angular scales ($\ell >
   100$).}
\begin{tabular}{|c|c|c|c|c|c|c|}
\hline
 &counts&$\gamma \gamma$&{{\rm counts}+$\gamma \gamma$}&$\eta$&$\gamma \gamma^l$ &{{\rm counts}+$\gamma \gamma^l$}\\
\hline
$\Delta \Omega_{\rm DE}$  &0.064   &0.0087   &0.0026    &0.094     &0.014    &0.0027\\
$\Delta \Omega_m h^2$     &0.20    &0.051    &0.0075    &0.023    &0.036    &0.0060\\
$\Delta \sigma_8$         &0.029   &0.012    &0.0032    &0.081     &0.022    &0.0031\\
$\Delta w_0$              &0.080   &0.081    &0.034     &0.35     &0.13     &0.032\\
$\Delta w_a$              &1.24    &0.29     &0.12      &0.18     &0.45     &0.10\\
$\Delta \Omega_b h^2$     &0.0010  &0.015    &0.0010    &1.0      &0.012    &0.00099\\
$\Delta n_s$              &0.040   &0.052    &0.019     &0.38     &0.028    &0.013\\
\hline
\end{tabular}
\end{table}

A related issue is that in scalar field models of dark
energy~(e.g.\cite{T&W97}), such as quintessence~\cite{CD&S98}, the
field clusters on large scales, while it remains smooth on small
scales. This is different from a cosmological constant, which remains
smooth on all scales, and the additional dark--energy fluctuations can
enhance the matter fluctuations on large scales. It is interesting to
ask whether this enhancement may be detectable through the
shear--shear correlations~\cite{DH01}.  We take, as an example, the
shear--shear auto power spectra ($C_\ell^{88}$) of the 8th. Compared
with other bins, this has the largest comoving radial distance, so
that for fixed $\ell$, it probes the largest comoving scales, and
should be most sensitive to the clustering effect of quintessence
field. We calculate $C_\ell^{88}$ for both our fiducial model (with a
cosmological constant) and a quintessence cosmological model with
$w_0=-0.5$, with all other parameters fixed. In the quintessence
model, we use KINKFAST with the choice for the transfer function that
includes dark energy perturbations.

The quintessence field affects the power spectrum through the
expansion rate, the growth rate and the enhancement of the matter
fluctuations on large scales, causing the deviations of these two
$C_\ell^{88}$ curves from each other.  To separate the clustering
effect from the other two effects, we artificially replace the
transfer function in the $w=-0.5$ quintessence cosmological model by
the one that excludes the effect of dark energy clustering (i.e. the
$w=-1$ transfer function), and calculate $C_\ell^{88}$ for the
quintessence cosmological model again.  We find that the difference
caused by dark energy clustering is unfortunately quite small, safely
within the error bars. A remaining issue is that the above treatment
only computes the fluctuations in the matter density induced by dark
energy clustering. On the other hand, the weak lensing signal is
sensitive to fluctuations in the total gravitational potential, which
has an additional contribution directly from the fluctuations in the
dark energy component.  However, we expect these two contributions to
be of similar order of magnitude. We conclude, in agreement with
ref.~\cite{DH01} that while the shear--shear correlations can tell the
quintessence field apart from a cosmological constant, the distinction
is made purely through the effect on the growth rate and expansion
rate, and the clustering of dark energy on large scales remains
undetectable.

\subsection{Shear Power Spectrum vs. Cluster Power Spectrum}

Once the galaxy clusters are detected, their spatial distribution,
characterized, e.g., by the cluster power spectrum ($P_c(k)$), readily
offers another constraint on cosmology.  Since earlier
works~\cite{M&M04,SW04} have studied the complementarity of the
cluster counts and their power spectrum, here we contrast the
$dN/dz+P_c(k)$ combination with the $dN/dz+C_\ell$ combination. To
perform this comparison, we divide the clusters according to their
redshift into 6 bins, each with size of $\Delta z=0.2$, except the
farthest one with size of $\Delta z=0.3$. For each bin, we follow Hu
\& Haiman~\cite{H&H03} and compute the cluster power spectrum over
$30\times 30$ $k$--space cells centered at $k_{\parallel},
k_{\perp}=0.005,0.010,...,0.15~{\rm Mpc}^{-1}$, where $k_{\parallel},
k_{\perp}$ are wave--numbers parallel and transverse to the line of
sight, respectively. The methods to obtain constraints on cosmological
parameters from the cluster power spectrum are the same as those
described in~\cite{SW04}. Our results are shown in
Table~\ref{tab:cps}, separately for the number counts (as before), the
power spectrum, and their combination.  We have simply summed the two
Fisher matrices, assuming that the two measurements are independent
and have no covariance.  This assumption was implicitly made in
previous works~\cite{SW04,M&M04}, but it could be justified by
arguments similar to those in the previous section.  The $5^{\rm th}$
column in the Table shows the complementarity parameter, defined
analogously to that in Table \ref{tab:main}.

A comparison of Table \ref{tab:main} and Table \ref{tab:cps} reveals
that shear--shear correlations give tighter constraints on
$\Omega_{\rm DE}, w_0, w_a$ than the cluster power spectrum, either by
itself, or in combination with the number counts.  In particular, the
combined constraints are a factor of $\sim 1.5-2$ better when the
shear--shear correlations are used. The shear--shear correlations
utilize structure information within the redshift range of [0, 3.2],
while cluster power spectrum survey utilizes only that in the redshift
range of [0.1, 1.4] (few clusters can be detected at redshifts beyond
this range, see ~\cite{SW04}). In Table~\ref{tab:4bin}, we show the
constraints as in Table~\ref{tab:main}, except that we have used only
the first four redshift bins for the shear--shear correlations, that
is, only the source galaxies with redshift [0,1.6] are considered.
The Table shows that the constraints would typically degrade by $\sim
30\%$ if the high--$z$ tail of galaxies were discarded.  We conclude
that about $\sim$half of the advantage of the shear--shear
correlations over the cluster power spectrum comes from this high--$z$
tail; the rest of the improvement is due to the fact that at low
redshift, the shear--shear correlations (with $\ell_{\rm max}=1000$
corresponding to $k\sim 1$ at $z\sim 0.2$), probe smaller spatial
scales than we used for the power spectrum ($k_{\rm max}= 0.15 {\rm
Mpc}^{-1}$).

\begin{table}
  \caption{\label{tab:cps} {\it This table replaces the shear--shear
correlations by the cluster power spectrum.}  Marginalized errors are
shown on cosmological parameters from cluster counts, cluster power
spectrum, and their combination. The parameter $\eta$, shown in the
$5^{\rm th}$ column, measures the synergy between the two observables,
as in Table 1.}
\begin{tabular}{|c|c|c|c|c|}
\hline
&counts&$P_c(k)$&{{\rm counts}+$P_c(k)$}&$\eta$\\
\hline
$\Delta \Omega_{\rm DE}$ &0.064   &0.0095   &0.0040    &0.18\\
$\Delta \Omega_m h^2$     &0.20    &0.027    &0.0047    &0.030\\
$\Delta \sigma_8$         &0.029   &0.020    &0.0050    &0.091\\
$\Delta w_0$              &0.080   &0.12     &0.057     &0.75\\
$\Delta w_a$              &1.24    &0.54     &0.23      &0.23\\
$\Delta \Omega_b h^2$     &0.0010  &0.0063   &0.00097   &0.97\\
$\Delta n_s$              &0.040   &0.046    &0.013     &0.19\\
\hline
\end{tabular}
\end{table}

\begin{table}
  \caption{\label{tab:4bin} {\it This table shows the effect of
excluding the highest redshift galaxies from the shear measurements.}
Marginalized errors are shown on the cosmological parameters, as in
Table~\ref{tab:main}, except that here we have used only the first
four redshift bins for the shear--shear correlations, that is, only
the source galaxies with redshift [0,1.6] are considered.}
\begin{tabular}{|c|c|c|c|c|c|c|}
\hline
 &counts&$\gamma \gamma$&{{\rm counts}+$\gamma \gamma$}&$\eta$&$\gamma \gamma^l$ &{{\rm counts}+$\gamma \gamma^l$}\\
\hline
$\Delta \Omega_{\rm DE}$  &0.064   &0.011   &0.0034    &0.11     &0.035    &0.0035\\
$\Delta \Omega_m h^2$     &0.20    &0.11    &0.0095    &0.0093    &0.065    &0.0070\\
$\Delta \sigma_8$         &0.029   &0.018   &0.0042    &0.076     &0.080    &0.0044\\
$\Delta w_0$              &0.080   &0.10    &0.046     &0.53     &0.23    &0.049\\
$\Delta w_a$              &1.24    &0.40    &0.17      &0.21     &1.18     &0.18\\
$\Delta \Omega_b h^2$     &0.0010  &0.032   &0.0010    &1.0      &0.020   &0.0010\\
$\Delta n_s$              &0.040   &0.12    &0.025     &0.43     &0.041    &0.014\\
\hline
\end{tabular}
\end{table}

\subsection{The Impact of Systematic Errors}

The above results (listed in Table~\ref{tab:main}) are encouraging,
and suggest that cluster counts will be a useful complement to the
shear--shear correlations, despite the fact that constraint from the
counts alone are weaker.  However, a general concern with this
conclusion is that systematic errors, which will inevitably degrade
constraints from individual observables, may additionally degrade
their synergy.  Here we briefly examine some aspects of this question.

First, the major concern with selecting clusters from their weak
lensing shear alone is that projection effects will produce false
detections (contamination) and cause real clusters to drop out of
the sample (incompleteness).~\cite{H&S05,H&T&Y03,W&VW&M02} In
principle, these effects can be modeled in ab--initio simulations,
but for a very large survey, such as LSST, the contamination and
completeness has to be quantified to a very stringent fractional
accuracy of $N_{\rm total}^{-1/2}\sim 2\times 10^{-3}$ in order not
to dominate Poisson errors.  Alternative approaches would be to
utilize other (optical or X--ray) data to improve the accuracy of
the selection function (see ref.~\cite{SW04} for more discussion).

Here we note that if we restrict our analysis to increasingly
high--$\sigma$ shear peaks, our results should become increasingly
realistic. This is for two reasons: (i) contamination and completeness
improve rapidly as the detection threshold (or cluster mass) is
increased,~\cite{H&S05,H&T&Y03,W&VW&M02} and (ii) these peaks are
rare, and therefore the required $N_{\rm total}^{-1/2}$ accuracy for
the selection function becomes less stringent, and easier to achieve
in simulations.  Here we simply examine the effect of increasing the
threshold $\kappa_{\rm th}$, to quantify whether a smaller cluster
sample, derived from higher shear peaks, is still useful. In
Table~\ref{tab:kth10}, we repeat our calculations from
Table~\ref{tab:main}, except that we replace the threshold
$\kappa_{\rm th}=4.5 \sigma_{\rm noise}$ by $\kappa_{\rm th}=10, 20,
30 \sigma_{\rm noise}$.  The table shows that the number of clusters
diminish rapidly: $N_{\rm total}=276,794\rightarrow 30,554\rightarrow
1,954\rightarrow 205$, respectively, as the threshold is increased.
On the other hand, despite this decrease, cluster counts remain useful
in tightening the constraints.  For example, the most massive $\sim
30,000$ clusters still improve dark energy constraints by a factor of
two relative to using the shear--shear correlations alone.  Even the
most massive $\approx 200-2,000$ clusters, which, by themselves, do
not offer interesting constraints on dark energy, still improve the
constraints when added to the shear--shear correlations.  This gives
us confidence that cluster counts will be a useful complement to the
shear--shear correlations, despite systematic errors in the weak
lensing cluster selection function.

\begin{table}
\caption{\label{tab:kth10} {\it This table examines raising the
shear--detection threshold.}  Marginalized errors are shown on
cosmological parameters, as in Table~\ref{tab:main}, except that we
adopt increasingly more stringent detection thresholds for the
convergence.  Note that cluster counts improve dark energy constraints
when added to the shear--shear correlations, even for exceedingly high
detection thresholds.}
\begin{tabular}{|c|c|c|c|c|}
\hline
 &counts&$\gamma \gamma$&{{\rm counts}+$\gamma \gamma$}&$\eta$\\
\hline
\multicolumn{5}{|c|}{$\kappa_{\rm th}=10 \sigma_{\rm noise}, N_{\rm total}=30,554$}\\
\hline
$\Delta \Omega_{\rm DE}$ &0.10   &0.0084   &0.0046    &0.30 \\
$\Delta \Omega_m h^2$     &0.35    &0.049    &0.0062    &0.017\\
$\Delta \sigma_8$         &0.14   &0.012    &0.0054    &0.21 \\
$\Delta w_0$              &0.28   &0.078    &0.048     &0.40 \\
$\Delta w_a$              &2.18    &0.28     &0.14     &0.26 \\
$\Delta \Omega_b h^2$     &0.0010  &0.014    &0.00099   &0.99 \\
$\Delta n_s$              &0.040   &0.050    &0.017     &0.30 \\
\hline
\multicolumn{5}{|c|}{$\kappa_{\rm th}=20 \sigma_{\rm noise}, N_{\rm total}=1,954$}\\
\hline
$\Delta \Omega_{\rm DE}$ &0.083   &0.0084   &0.0065   &0.61 \\
$\Delta \Omega_m h^2$     &0.71   &0.049    &0.0063    &0.017\\
$\Delta \sigma_8$         &0.33   &0.012    &0.0084    &0.51 \\
$\Delta w_0$              &2.78  &0.078    &0.060     &0.59 \\
$\Delta w_a$              &6.96    &0.28     &0.19     &0.45 \\
$\Delta \Omega_b h^2$     &0.0010  &0.014    &0.00099   &0.99 \\
$\Delta n_s$              &0.040   &0.050    &0.017     &0.31 \\
\hline
\multicolumn{5}{|c|}{$\kappa_{\rm th}=30 \sigma_{\rm noise}, N_{\rm total}=205$}\\
\hline
$\Delta \Omega_{\rm DE}$ &0.52   &0.0084   &0.0075    &0.81 \\
$\Delta \Omega_m h^2$     &1.21    &0.049    &0.0064    &0.018\\
$\Delta \sigma_8$         &0.44   &0.012    &0.010    &0.76 \\
$\Delta w_0$              &12.3   &0.078    &0.067     &0.74 \\
$\Delta w_a$              &39.4    &0.28     &0.22     &0.65 \\
$\Delta \Omega_b h^2$     &0.0010  &0.014    &0.00099   &0.99 \\
$\Delta n_s$              &0.040   &0.050    &0.018     &0.33 \\
\hline
\end{tabular}
\end{table}

Another significant concern is that clusters are not spherical
structures, and, when viewed from different directions, will produce a
different shear.  Simulations suggest that this can cause an
irreducible scatter, and possibly a bias, in the relation between halo
mass and shear~\cite{RC97,M&W&L01,C&G&M04} In addition, fluctuations
caused by large--scale structure along the line of sight will
introduce a scatter.  While once again these effects can be studied in
ab--initio simulations, and may be correctable
statistically~\cite{SD04} or by identifying foreground lensing
galaxies and directly subtracting their lensing effect~\cite{dP&W05},
the accuracy to which the magnitude and shape of the unknown scatter
will be reduced is not yet clear.  Here we perform a simple exercise,
and model the probability distribution $p(\kappa | M,z)$for a dark
matter halo with fixed mass $M$ at redshift $z$ to produce a smoothed
convergence $\kappa$ to be given by a Gaussian,
\begin{equation}
p(\kappa | M,z)=
\frac{1}{\sqrt{2\pi}\sigma_{\kappa}}
\exp\left[-\frac{(\kappa-\kappa_{\rm NFW})^2}{2\sigma_{\kappa}^2}\right],
\end{equation}
where $\kappa_{\rm NFW}$ is calculated by assuming an NFW density
profile for the dark matter halo as described in \S.~\ref{sec:II}
above, and we assume $\sigma_{\kappa}=\epsilon \kappa_{\rm NFW}$.  We
assume that the value of $\kappa$ at fixed mass is known ab--initio to
within $\sim 30\%$, i.e we adopt the fiducial value of
$\epsilon=0.3$. The probability of detecting this dark matter halo by
setting a detection threshold of $\kappa_{\rm th}$ is then
\begin{equation}
P(M,z)=\frac{1}{2}
{\rm erfc}\left[\frac{\kappa_{\rm th}-\kappa_{\rm NFW}}{\sqrt{2}\sigma_{\kappa}}\right],
\end{equation}
and equation~(\ref{eqn:Ni}) is modified to
\begin{equation}
N_i=\Delta \Omega \Delta
z\frac{d^2V}{dzd\Omega}(z_i)\int_{0}^{\infty}\frac{dn}{dM}(M,z_i)P(M,z_i)dM.
\end{equation}
First, this assumed fiducial scatter increases the number of the total
detected clusters (from 276,794 to 305,385).  We then recompute the
constraints, letting $\epsilon$ to be an additional free parameter,
adopting a weak prior of $\Delta\epsilon=0.3$.  The estimated
uncertainties on the cosmological parameters from number counts after
marginalizing over $\epsilon$, and its combination with shear-shear
correlations, are listed in Table~\ref{tab:asphe}.  A comparison with
Table~\ref{tab:main} reveals that the cluster--count constraints
degrade somewhat due to this uncertain scatter, the combined
constraint degrade very little.  Therefore, we conclude that as long
as the $\kappa-M$ relation can be characterized ab--initio to within
$\sim 30\%$, we expect our results to remain realistic.
\begin{table}
\caption{\label{tab:asphe} {\it This table examines the effect of
scatter in the mass--convergence relation.} Marginalized errors are
shown on cosmological parameters, as in Table~\ref{tab:main}, except
that we allow for an additional free parameter, $\epsilon$,
representing a scatter between cluster mass and the convergence it
produces.  We assume a Gaussian distribution for this scatter, and
adopt a prior of $\Delta \epsilon=0.3$ when the constraints from the
number counts are marginalized over $\epsilon$.}
\begin{tabular}{|c|c|c|c|c|c|c|}
\hline
 &counts&$\gamma \gamma$&{{\rm counts}+$\gamma \gamma$}&$\eta$&$\gamma \gamma^l$ &{{\rm counts}+$\gamma \gamma^l$}\\
\hline
$\Delta \Omega_{\rm DE}$ &0.11   &0.0084   &0.0026    &0.096   &0.013     &0.0028\\
$\Delta \Omega_m h^2$     &0.35    &0.049    &0.0061    &0.016   &0.034     &0.0053\\
$\Delta \sigma_8$         &0.032   &0.012    &0.0034    &0.096   &0.021     &0.0052\\
$\Delta w_0$              &0.077   &0.078    &0.032     &0.35    &0.12      &0.035\\
$\Delta w_a$              &1.32    &0.28     &0.11      &0.17    &0.42      &0.14\\
$\Delta \Omega_b h^2$     &0.0010  &0.014    &0.00099   &0.99    &0.010     &0.00099\\
$\Delta n_s$              &0.040   &0.050    &0.016     &0.28    &0.027     &0.012\\
\hline
\end{tabular}
\end{table}

Another issue is whether photometric redshift accuracies will limit
the constraints quoted here.  In the case of shear tomography, the
impact of redshift uncertainties has been considered in detail in
refs~\cite{M&H&H06,H&T&B&J06}.  While our redshift bins are relatively
wide, calibration of the photometric errors to the accuracy required
to avoid degrading dark energy parameter constraints will likely
require a large spectroscopic follow--up program.  We refer the reader
to refs~\cite{M&H&H06,H&T&B&J06} for detailed treatments.  In the case
of cluster counts, we used 26 redshift bins in our analysis,
effectively requiring that we know cluster redshifts to within $\Delta
z \approx 0.05$. This is comparable to expected photometric redshift
errors, and should be feasible to achieve for most of the clusters,
for which a secure identification of cluster membership can be made
for a few galaxies.  On the other hand, the use of 26 bins is not
actually required -- one expects that fewer bins are sufficient, since
the cluster abundance varies relatively smoothly with redshift.

\begin{figure}[t!]
 \resizebox{90mm}{!}{\includegraphics{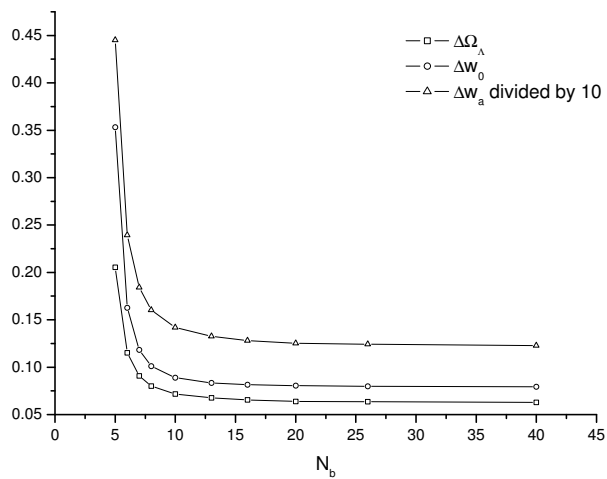}}
 \caption{\label{fig:5} The marginalized errors are shown on dark
 energy parameters from the cluster number counts, as a function of
 the number of redshift bins ($N_b$) used in the analysis.  The bins
 are assumed to be equally spaced in redshift.  The flatness of the
 curves for $N_b\gsim 10$ shows that the full information content of
 the abundance evolution can be extracted with rather modest cluster
 redshift accuracies of $\Delta z \approx 0.15$.  }
\end{figure}

To address this issue, in Figure~\ref{fig:5} we show the marginalized
errors on the dark energy parameters from the number counts, as a
function of the number of redshift bins $N_b$.  The flatness of the
curves on the figure for $N_b\gsim 10$ shows that the full information
content of the abundance evolution can be extracted with rather modest
cluster redshift accuracies of $\Delta z \approx 0.15$.  This
accuracy, however, is still a factor of $\sim 3$ more stringent than
the r.m.s.  redshift errors expected to be available from tomography
alone~\cite{H&S05}.  Figure~\ref{fig:5} shows that with tomographic
redshifts alone (i.e. with only $N_b\sim3$ redshift bins), there would
be no interesting constraints on dark energy parameters. Hence, to
realize the full potential of the survey, it will be important to
securely identify member galaxies in the low--mass clusters and groups
at the detection threshold.

\section{Conclusions}
\label{sec:V}

In this paper, we studied the possibility of improving constrains on
dark energy properties by combining two observables -- the number
counts of detected galaxy clusters and large angular scale tomographic
shear--shear correlations -- that will both be automatically available
in a future large weak lensing survey.  We showed that the covariance
between these two observables is small, and argued that they can
therefore be regarded as independent constraints on dark energy
parameters.  We used the Fisher matrix formalism to forecast the
expected statistical errors from either observable, and from their
combination.  We found that combining the two observables results in
an improvement on dark energy parameter uncertainties by a factor of
$2-25$, relative to using either observable alone.  We have argued
that this conclusion may survive in the face of systematic errors.
Our results also suggest that neither observable may exhaust the full
information content of a non--linear weak lensing map.\\

\begin{acknowledgments}

  We thank Greg Bryan and Lam Hui for helpful discussions and Sheng
  Wang and Bhuvnesh Jain for useful comments on the manuscript.  This
  work was supported in part by NSF grant AST-05-07161 (to Z.H.)  and
  by the Initiatives in Science and Engineering (ISE) program at
  Columbia University.

\end{acknowledgments}


\onecolumngrid
\appendix
\begin{center}
  {\bf APPENDIX}
\end{center}

In this Appendix, we give the details of our calculation of the
covariance between the cluster counts and the expansion coefficients
of the E modes of the shear maps, i.e. equation~(\ref{eqn:cov-ns}).

\subsection{Sample Variance of Number Counts}
Following ~\cite{L&H04}, we model the number counts of galaxy clusters
in each redshift bin with independent Poisson distributions, whose
mean are drawn from correlated Gaussian distributions. Given a
cosmological model, the probability density of drawing $\{N_i;
i=1,...n, n=26\}$ clusters, is given by
\begin{equation}
P({\mathbf{N}}|\mathbf{\bar N},\mathbf{S}^{\rm counts})=\int d^nM
\left[\prod_{i=1}^{n} P(N_i|M_i)\right]G(\mathbf{M}|\mathbf{\bar
N},\mathbf{S}^{\rm counts})
\end{equation}
where $P(N_i|M_i)$ is the normalized Poisson distribution for $N_i$
with a mean $M_i$, and $G(\mathbf{M}|\mathbf{\bar N},\mathbf{S}^{\rm
counts})$ is the normalized multi--variate Gaussian probability
distribution for the $M_i$ with mean $\mathbf{\bar N}$. Here
$\mathbf{N}$, $\mathbf{\bar{N}}$ and $\mathbf{M}$ are
26--dimensional column vectors, representing the counts in the 26
redshift bins, and $\mathbf{S}^{\rm counts}$ is the $26\times 26$
covariance matrix. The integral represents averaging over a
fluctuating mean $\mathbf{M}$. These fluctuations are due to
large--scale fluctuations in the underlying matter density field.
Under the assumption that clusters trace the matter density field
with a linear bias, we have
\begin{equation}
M_i-\bar{N_i}=V_i \int d^3 x
W_i(\vec{x}(\chi,\hat{n}))b(\chi)\delta(\vec{x}(\chi,\hat{n}))\overline{n}(\chi),
\label{eq:countexpansion}
\end{equation}
Where we use the comoving radial distance $\chi$ as the time
coordinate for the time--dependent quantities, $V_i$ is the comoving
volume for the $i^{th}$ redshift bin of cluster counts, $b(\chi)$ is
the cluster--averaged linear bias~\cite{SW04,S&T99},
$\delta(\vec{x}(\chi,\hat{n}))$ is the density contrast field of
matter, and $W_i(\vec{x}(\chi,\hat{n}))$ is the normalized survey
window function, which we model to be part of a spherical shell.
With $W_i(\vec{x}(\chi,\hat{n}))=R_i(\chi)\Theta(\hat{n})$, we have
\begin{equation}
R_i(\chi)=\begin{cases}3 \left[(\chi_{max}^i)^3-(\chi_{min}^i)^3
\right]^{-1}, & \chi \in[\chi_{min}^i,\chi_{min}^i] \\ 0, &\rm
otherwise
\end{cases}
\end{equation}
\begin{equation}
\Theta(\hat{n}(\theta,\varphi))=\begin{cases}
\left[2\pi(1-\cos{\theta_s})\right]^{-1}, & \theta \in[0,\theta_s], \varphi \in[0,2\pi)\\
0, &\rm otherwise
\end{cases}
\end{equation}
Where, $\theta_s$ is the angular size of the survey region, (in our
case, for 18,000 deg$^2$ centered on the pole at $\theta=0$,
$\theta_s=\arccos[1-\frac{5}{18\pi}]$), and $\Delta \chi^i$
($=\chi_{\rm max}^i-\chi_{\rm min}^i$) is the comoving radial extent
of this $i^{\rm th}$ redshift bin.  $\overline{n}(\chi)$ is the
expected comoving number density of the detectable clusters. It is
related to $\bar{N_i}$ by
\begin{equation}
\bar{N_i}=V_i \int d^3x
W_i(\vec{x}(\chi,\hat{n}))\overline{n}(\chi). \label{eqn:Ni-exact}
\end{equation}
In the limit $\Delta \chi^i \rightarrow 0$, we have
\begin{equation}
R_i(\chi)\rightarrow \frac{1}{\chi_{z_i}^2}\delta(\chi-\chi_{z_i}),
\end{equation}
equation~(\ref{eqn:Ni-exact}) reduces to equation~(\ref{eqn:Ni}).

The elements of the sample covariance matrix, $\mathbf{S}^{\rm
counts}$, can be calculated by~\cite{H&K03}
\begin{multline}
S^{\rm counts}_{ij}\equiv\langle(M_i-\bar{N_i})(M_j-\bar{N_j})\rangle=
\bar{N_i}\bar{N_j}
b(z_i)b(z_j)D(z_i)D(z_j)
\sum_\ell 4\pi \tilde{\Theta_\ell}^2
\int \frac{dk}{k}
\tilde{R}_{i\ell}(k)\tilde{R}_{j\ell}(k)\Delta^2(k,z=0).
\end{multline}
Here $D(z)$ is the growth factor of mass fluctuations, normalized to
unity today, $\Delta^2(k,z=0)$ is the present--day variance per $\ln
k$ for the matter fluctuations, and $\tilde{\Theta_\ell},
\tilde{R}_{i\ell}(k)$ are quantities related to the Fourier
transform of the survey window function, given by
\begin{equation}
\tilde{\Theta_\ell}=\begin{cases}\sqrt{\frac{1}{4\pi}},& \ell=0 \\
\sqrt{\frac{2\ell+1}{4\pi}}\frac{(1+x)}{\ell(\ell+1)}
\frac{d}{dx}P_\ell(x)|_{x=\cos\theta_s},& \ell\geq 1
\end{cases}
\end{equation}
and
\begin{equation}
\tilde{R}_{i\ell}(k)\simeq \frac{1}{\Delta\chi^i}\int_{\chi_{\rm
min}^i}^{\chi_{\rm max}^i}d\chi j_\ell(k\chi),
\end{equation}
with $P_\ell(x)$ the $\ell^{\rm th}$ order Legendre polynomials, and
$j_\ell(k\chi)$ the $\ell^{\rm th}$ order spherical Bessel
functions.

\subsection{Sample Variance of Shear}

Under the Gaussian assumption for the E mode of the shear field, the
probability of drawing a set of spherical expansion coefficients
$a_{\ell m}^b$ in a given cosmological model is
\begin{equation}
P(\mathbf{a}|\mathbf{C}^{\gamma\gamma})=G(\mathbf{a}|\mathbf{0},\mathbf{C}^{\gamma\gamma})
\end{equation}
where $\mathbf{a}$ is column vector of all the $a_{\ell m}^b$ (for our
range of $41\leq \ell \leq 1000$ and with 8 redshift bins, this vector
has dimension $8\times\Sigma_{41}^{1000}(2\ell+1)=8,002,560$),
$\mathbf{0}$ is a column zero vector of the same dimension, and
$\mathbf{C}^{\gamma\gamma}=\mathbf{S}^{\gamma\gamma}+\mathbf{N}^{\gamma\gamma}$
is the covariance matrix for the $a_{\ell m}^b$, and
$\mathbf{S}^{\gamma\gamma}$ and $\mathbf{N}^{\gamma\gamma}$ are as
given in Sec.~\ref{sec:II B}.  Note that both
$\mathbf{S}^{\gamma\gamma}$ for the shear field and $\mathbf{S}^{\rm
counts}$ for the cluster number counts come from fluctuations in the
matter distribution. Since the shear field and the clusters are
detected in the same realization of the matter distribution, they can
not be independent.

The convergence (or E mode of the shear field) signal from the source
galaxies in redshift bin $b$ is calculated by
\begin{equation}
\kappa^b(\hat{n})= \frac{1}{2c^2}\int_0^{\chi_{\infty}}d\chi
(\bigtriangledown^2-\bigtriangledown_{\chi}^2)
\Phi(\vec{x}(\chi,\hat{n}))W^b(\chi)
\end{equation}
where $\Phi(\vec{x}(\chi,\hat{n}))$ is the gravitational potential
field. As discussed in~\cite{J&S&W00}, the integral of the term that
contains $\bigtriangledown_{\chi}^2 \Phi$ is much smaller than that
containing $\bigtriangledown^2 \Phi$ everywhere except on the
largest angular scales. We thus neglect this term in the following
calculation. Expanding $\kappa^b(\hat{n})$ in spherical harmonics,
we get
\begin{equation}
a_{\ell m}^b=\int d\Omega \kappa^b(\hat{n}) Y_{\ell
m}^*(\hat{n})+\rm noise.
\end{equation}
The Poisson equation reads as
\begin{equation}
\bigtriangledown^2\Phi(\vec{x}(\chi,\hat{n}))=4\pi G
a^2(\chi)\rho_m(\chi)\delta(\vec{x}(\chi,\hat{n})),
\end{equation}
with $\delta(\vec{x}(\chi,\hat{n}))$ in this equation representing the
(non--linear) matter density contrast. Using this equation, we find
\begin{equation}
a_{\ell m}^b=\frac{2\pi G}{c^2}\int d \chi
a^2(\chi)\rho_m(\chi)W^b(\chi)\int d\Omega Y_{\ell
m}^*(\hat{n})\delta(\vec{x}(\chi,\hat{n}))+\rm noise.
\label{eq:shearexpansion}
\end{equation}

The sample variance $\bf{S}^{\gamma\gamma}$ and full covariance
$\bf{C}^{\gamma\gamma}$ can then be computed as the ensemble average
$\langle a_{\ell m}^b a_{\ell' m'}^{b'*} \rangle$ with or without the
noise term (the latter leads to equation~\ref{eq:Fishershear}).

\subsection{Covariance Between Number Counts and Shear}

The correlation between $(M_i-\bar{N_i})$ and $a_{\ell m}^b$ is
nonzero because they depend on the same density field. For clarity,
from now on, we add a prime to the coordinates relating to the
calculation of $a_{\ell m}^b$. The correlation between the two density
contrast fields is given by the two--point function
\begin{equation}
\langle\delta(\vec{x}(\chi,\hat{n}))\delta(\vec{x'}(\chi',\hat{n}'))
\rangle=D(\chi)D(\chi')\int \frac{d^3k}{(2\pi)^3} e^{i \vec{k}\cdot
(\vec{x}-\vec{x'})} P(k,\chi_p=0 ).
\end{equation}

The covariance of $(M_i-\bar{N_i})$ and $a_{\ell m}^b$ is then given
by
\begin{multline}
\langle(M_i-N_i)a_{\ell m}^b\rangle=\frac{2\pi G}{c^2}V_i \int
\chi^2 d \chi R_i(\chi) b(\chi) \overline{n}(\chi) D(\chi) \int
d\chi' a^2(\chi')\rho_m(\chi')W^b(\chi')D(\chi') \\\int \frac{k^2
dk}{(2\pi)^3} P(k,\chi_p=0 ) \int d \Omega_k \int d \Omega' e^{-i
k\chi' \hat{k} \cdot \hat{n'}} Y_{\ell m}^*(\hat{n'}) \int d \Omega
e^{i k\chi \hat{k} \cdot \hat{n}} \Theta(\hat{n}).
\end{multline}
Note noise from intrinsic ellipticity of galaxies does not correlate
with this fluctuation of cluster counts, so the noise term drops
here. The integral over $\Omega$ gives
\begin{equation}
I_{\Omega}=\sum_{\ell'=0}^{\infty}\sum_{m=-\ell'}^{\ell'} 4\pi
i^{\ell'} j_{\ell'}(k\chi)Y_{\ell'
m'}(\hat{k})\tilde{\Theta}_{\ell'}\delta_{m'0},
\end{equation}
and the integral over $\Omega '$ gives
\begin{equation}
I_{\Omega'}=4\pi (-i)^{\ell}Y_{\ell m}^*(\hat{k})j_{\ell}(k\chi').
\end{equation}

We are now ready to calculate the integral over $\Omega_k$, which
gives
\begin{equation} I_{\Omega_k}=\delta_{m0}(4\pi)^2
\tilde{\Theta}_{\ell}j_{\ell}(k\chi)j_{\ell}(k\chi').
\end{equation}
We next use the Limber approximation when integrating over
$k$. Assuming that $P(k,\chi_p=0)$ is a slowly varying function
compared to the $j_{\ell}(k\chi)$, we have
\begin{equation}
I_k\simeq \delta_{m0}
\tilde{\Theta}_{\ell}\frac{1}{\chi^2}\delta(\chi-\chi')P(k=\frac{\ell}{\chi},\chi_p=0),
\end{equation}
where we have used the orthogonality property
\begin{equation}
\int_0^{\infty} k^2dk
j_{\ell}(k\chi)j_{\ell}(k\chi')=\frac{\pi}{2\chi^2}\delta(\chi-\chi')
\end{equation}
for $\ell>0$. The integral over $\chi'$ is easy to perform, and we finally
have
\begin{equation}
\langle(M_i-N_i)a_{\ell m}^b\rangle=\frac{2\pi G}{c^2}
\delta_{m0}V_i \tilde{\Theta}_{\ell}\int
 d \chi R_i(\chi) b(\chi) \overline{n}(\chi)
a^2(\chi)\rho_m(\chi)W^b(\chi)P(k=\frac{\ell}{\chi},\chi).
\end{equation}
recalling the evolution of the matter density,
\begin{equation}
\rho_m(\chi)=\Omega_m \frac{3H_0^2}{8\pi G}\frac{1}{a(\chi)^3},
\end{equation}
and of the power spectrum,
\begin{equation}
\Delta^2(k,\chi)=\frac{k^3}{2\pi^2}P(k,\chi),
\end{equation}
by taking the limit of $\Delta \chi^i \rightarrow 0$, we obtain the
final result.
\begin{equation}
\langle(M_i-N_i)a_{\ell
m}^b\rangle=\delta_{m0}\frac{3\pi^2}{2}\frac{\Omega_m
H_0^2}{c^2a(z_i)} \frac{\chi_{z_i}}{\ell^3} \tilde{\Theta_\ell}N_i
b(z_i)W^b(z_i)\Delta^2(k=\frac{\ell}{\chi_{z_i}},z_i).
\end{equation}

This result is quoted above in equation~(\ref{eqn:cov-ns}), and the
corresponding cross--correlation coefficient is evaluated explicitly
in one example in Figure~\ref{fig:4a}.

Given the above result on the cross correlation, we can write down the
expression for the joint probability distribution for simultaneously
drawing a set of $N_i$ and $a_{\ell,m}^b$,
\begin{equation}
P(\mathbf{N},\mathbf{a}|\mathbf{\bar{N}},\mathbf{C_{tot}})= \int d^n
M \left[\prod_{i=1}^{n}P(N_i|M_i)\right]
G(\mathbf{T}|\mathbf{\overline{T}},\mathbf{C_{tot}})
\end{equation}
where $\mathbf{T}$ is a column vector that combines $\mathbf{M}$ and
$\mathbf{a}$; $\mathbf{\overline{T}}$ is the column vector containing
the mean values $\mathbf{\bar{N}}$ and $\mathbf{0}$; and
$\mathbf{C_{\rm tot}}$ is the total covariance matrix, consisting of
four blocks,
\begin{equation}
\mathbf{C_{\rm tot}}=\left( \begin{matrix} \mathbf{S}^{\rm counts}& \mathbf{C}^{\rm cross}\\
(\mathbf{C}^{\rm cross})^{\rm T}& \mathbf{C}^{\gamma\gamma}
\end{matrix}\right)
\end{equation}
Note that here $\mathbf{C}^{\gamma\gamma}$ is the full
shear--covariance matrix, including the shear noise (in practice, we
found shear noise to have only a small effect on the results).  When
the $\xi_{i,b \ell m}$ are small, we can write the multi--variate
Gaussian as a product of two independent Gaussians,
\begin{equation}
G(\mathbf{T}|\mathbf{\overline{T}},\mathbf{C_{tot}})=
G(\mathbf{M}|\mathbf{\bar{N}},\mathbf{S}^{\rm counts})
G(\mathbf{a}|\mathbf{0},\mathbf{C}^{\gamma\gamma})+o(\xi),
\end{equation}
and the joint probability becomes separable,
\begin{equation}
P(\mathbf{N},\mathbf{a}|\mathbf{\bar{N}},\mathbf{C_{tot}})=
P(\mathbf{N}|\mathbf{\bar{N}},\mathbf{S}^{\rm counts}) P(\mathbf{a}|\mathbf{C}^{\gamma\gamma})+o(\xi),
\end{equation}
justifying the assumption that cluster number counts and shear-shear
correlations can be treated as independent cosmological probes.

\end{document}